\pdfoutput=1
\documentclass[conference]{IEEEtran}
\IEEEoverridecommandlockouts

\usepackage{cite}
\usepackage{amsmath,amssymb,amsfonts}
\usepackage{algorithmic}
\usepackage{graphicx}
\usepackage{textcomp}
\usepackage{xcolor}
\usepackage{booktabs}
\usepackage[protrusion=false]{microtype}
\usepackage{balance}
\usepackage{tikz}
\usetikzlibrary{positioning,arrows.meta,fit,backgrounds}

\def\BibTeX{{\rm B\kern-.05em{\sc i\kern-.025em b}\kern-.08em
T\kern-.1667em\lower.7ex\hbox{E}\kern-.125emX}}

\begin{document}
\bstctlcite{IEEEexample:BSTcontrol}

\title{Cross-Site Transfer and Spatial Forgetting\\in OTA-Trained Neural Receivers\thanks{This work has been submitted to the IEEE for possible publication. Copyright may be transferred without notice, after which this version may no longer be accessible.}}

\author{\IEEEauthorblockN{Riku Luostari}
\IEEEauthorblockA{\textit{Nokia} \\
Espoo, Finland \\
riku.luostari@nokia.com}
\and
\IEEEauthorblockN{Dani Korpi}
\IEEEauthorblockA{\textit{Nokia Bell Labs} \\
Espoo, Finland \\
dani.korpi@nokia-bell-labs.com}
\and
\IEEEauthorblockN{Harri Holma}
\IEEEauthorblockA{\textit{Nokia} \\
Espoo, Finland \\
harri.holma@nokia.com}
\and
\IEEEauthorblockN{Olav Tirkkonen}
\IEEEauthorblockA{\textit{Aalto University} \\
Espoo, Finland \\
olav.tirkkonen@aalto.fi}
}

\maketitle

\begin{abstract} Most published neural-receiver studies train on synthetic channel models, although real over-the-air measurements can improve performance. We examine how much of this benefit transfers from other sites to a distinctly different deployment site and what adaptation to that site costs elsewhere. The study uses a dataset from a 5.88 GHz measurement campaign covering eight sites and four environment types. An industrial site measured entirely from a vehicle is withheld as a deliberate transfer stress test. One receiver architecture is trained under four regimes: simulation alone; seven other sites; all sites jointly; and other-site training followed by deployment-site finetuning. The last two use identical measurements through different optimization protocols. Across five independently trained seeds, sensitivity gains are evaluated at uncoded bit error rate (BER) targets of \(10^{-1}\) and \(10^{-2}\), representative of moderate- and higher-rate coded links, respectively. On the deployment site, training with other-site data rather than simulation improves sensitivity by 0.3~dB at \(10^{-1}\) and 0.1~dB at \(10^{-2}\). Adding deployment-site data contributes less than 0.1~dB at \(10^{-1}\); at \(10^{-2}\), total gains over simulation reach 0.4~dB with joint training and 0.6~dB with finetuning. Joint training shows no measurable aggregate loss on the other-site test set, whereas finetuning yields the larger deployment-site gain but reduces other-site sensitivity by about 0.1~dB at \(10^{-2}\), indicating spatial forgetting.

\end{abstract}


\section{Introduction}

Deep learning replaces physical-layer blocks with learned functions, from channel estimation and equalization to receivers that map received signals directly to bit log-likelihood ratios \cite{oshea_introduction_2017, honkala_deeprx_2021,pihlajasalo_hybriddeeprx_2021,cammerer_neural_2023, luostari_input-correlated_2026}. Although foundational studies relied largely on synthetic channels, receivers trained on sufficiently randomized simulations can generalize over the air (OTA), and increasing the measured-data fraction improved an end-to-end receiver monotonically \cite{luostari_adapting_2025, luostari_input-correlated_2026}. This motivates a practical question: how does the value of OTA data depend on where it is collected? The degree of cross-site transfer determines whether a shared campaign is sufficient or deployment-site measurements provide worthwhile additional gain.

\begin{figure}[t]
\centering
\includegraphics[width=\columnwidth]{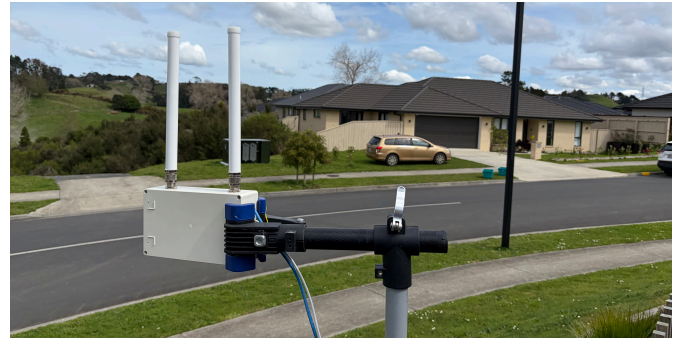}
\caption{The receiver unit with its antennas (Section~\ref{sec:measurement}).}
\label{fig:hardware}
\end{figure}

Transferability generally declines as source and target domains diverge \cite{yosinski_how_2014}. In a separate research line, Baytekin et al. established site-specific finetuning on real fifth-generation (5G) New Radio (NR) measurements and demonstrated transfer between deployment environments \cite{baytekin_site-specific_2026}. Wiesmayr et al. extended this work to additional receiver architectures, dual-layer transmission, and campaigns more than six months apart \cite{wiesmayr_impact_2026}. We extend this line of work to pretraining on OTA data from multiple sites, and compare joint training with finetuning on identical measurements. We therefore ask how far OTA data from seven sites transfers to a distinctly different deployment site and what adaptation to that site gives up elsewhere, a spatial analog of forgetting in continual learning \cite{kirkpatrick_overcoming_2017}.

We study these questions using the dataset from the 5.88~GHz measurement campaign introduced in \cite{luostari_input-correlated_2026}. It spans eight sites in urban, suburban, rural, and industrial environments. Both the transmitter and the receiver are software-defined radios (SDRs), forming an end-to-end link with a configurable radio interface. This configurability enables 6G physical-layer research and experimentation, including designs beyond existing specifications. Fig.~\ref{fig:hardware} shows the receiver unit. We hold out the most dissimilar site, an industrial area of scattered steel halls measured entirely from a vehicle at up to 15~m/s, as a deliberate transfer stress test. One receiver architecture is trained under four regimes: simulation alone; seven other sites; those sites jointly with 1,000 deployment-site measurements; and the other-site model finetuned on the same measurements.

\textbf{Contributions.} \emph{(i)~Cross-site transfer:} we quantify many-to-one transfer from seven source sites to a deliberately different deployment site, finding that other-site data adds the larger sensitivity increment at a moderate-rate operating point and deployment-site data at a higher-rate one. \emph{(ii)~Data-matched protocol comparison:} other-site versus all-site measures what deployment-site data adds; all-site versus finetuned compares joint training with finetuning on identical examples. \emph{(iii)~Spatial forgetting:} we quantify the reverse effect of deployment-site adaptation on the seven source sites; joint training shows no measurable aggregate loss on the other-site test set, while finetuning produces measurable forgetting but still outperforms simulation-only training at the other sites by about 0.4~dB. Five paired seeds quantify run-to-run variation in the reported differences.

\section{System Model}

We adopt the single-input single-output (SISO) orthogonal frequency-division multiplexing (OFDM) signal model and numerology of \cite{luostari_input-correlated_2026}, which builds on the OFDM chain of \cite{luostari_adapting_2025} (Table~\ref{tab:numerology}). With \(N_t = N_r = 1\), the received signal at subcarrier \(j\) and symbol \(i\) is
\begin{equation} y_{ij} = H_{ij} x_{ij} + n_{ij}, \end{equation}
where \(x_{ij}\) is the transmitted symbol, \(H_{ij}\) the complex channel gain, and \(n_{ij}\) additive white Gaussian noise (AWGN) with variance \(\sigma_n^2\). Each slot spans \(N_s = 14\) OFDM symbols and uses 16-ary quadrature amplitude modulation (16QAM). The fast Fourier transform (FFT) size is \(N_\mathrm{FFT}\); \(N_\mathrm{eff}\) subcarriers carry data or pilots after excluding guard bands and the direct-current (DC) subcarrier, with pilots at known resource-element indices \(\mathcal{P} \subset \{1, \dots, N_s\} \times \{1, \dots, N_\mathrm{eff}\}\). Bold uppercase symbols denote the per-slot time-frequency matrices (e.g., \(\mathbf{Y}, \mathbf{H}, \mathbf{X} \in \mathbb{C}^{N_s \times N_\mathrm{eff}}\)), with the \((i,j)\) subscript indexing an element.

\begin{table}[htbp]
\caption{OFDM system parameters, unchanged from \cite{luostari_input-correlated_2026}.}
\label{tab:numerology}
\centering
\footnotesize
\renewcommand{\arraystretch}{0.75}
\begin{tabular*}{\columnwidth}{@{\extracolsep{\fill}}lr@{}}
\toprule
\textbf{Parameter} & \textbf{Value} \\
\midrule
Carrier frequency & 5.88~GHz \\
Subcarrier spacing & 15~kHz \\
FFT size / effective subcarriers & 256 / 192 \\
OFDM symbols per slot & 14 \\
Modulation & 16QAM \\
\bottomrule
\end{tabular*}
\renewcommand{\arraystretch}{1}
\end{table}

The classical receiver against which every result is reported forms the least-squares (LS) estimate at the pilot positions,
\begin{equation}
\hat{H}^{\mathrm{LS}}_{ij} = \frac{y_{ij}}{x_{ij}} = H_{ij} + \frac{n_{ij}}{x_{ij}},
\quad (i,j)\in\mathcal{P}.
\end{equation}
The receiver fills the remaining resource elements by bilinear interpolation. Zero-forcing (ZF) equalization then divides out the interpolated estimate \(\hat{H}_{ij}\),
\begin{equation} \hat{x}_{ij} = \frac{y_{ij}}{\hat{H}_{ij}}, \end{equation}
after which the equalized symbols are demapped to per-bit log-likelihood ratios. We refer to this baseline as the LS receiver, or LS RX, chosen for its simplicity and mathematical tractability: every stage has a closed form and nothing in it is fitted to data, so it is identical across arms and sites. A linear minimum mean square error (LMMSE) estimator would provide a stronger baseline, but requires an assumed channel covariance; LS avoids this additional modeling choice.

\section{Methods}

\subsection{Hybrid CM DeepRx}
\label{sec:deeprx}

All experiments use Hybrid CM DeepRx, where CM denotes ConvMixer. We chose it for fast and consistent convergence across seeds rather than for peak sensitivity, making it well suited for investigating training regimes. It is derived from CM DeepRx, a ConvMixer realization of DeepRx \cite{luostari_input-correlated_2026,honkala_deeprx_2021,trockman_patches_2022}. It maps the received grid \(\mathbf{Y}\) and known pilots directly to per-bit log-likelihood ratios (LLRs) without forming an explicit channel estimate. Its backbone comprises six ConvMixer blocks with hidden dimension \(d = 256\), patch size one, asymmetric depthwise kernels, and an output convolution producing \(\log_2 M\) LLRs per resource element, where \(M\) is the modulation order. Relative to CM DeepRx, we widen the frequency kernel from \(K_f = 5\) to \(K_f = 15\), retain \(K_t = 3\), and move the first block's mixing to the delay domain.

The modified block adapts the hybrid idea of \cite{pihlajasalo_hybriddeeprx_2021}: an inverse discrete Fourier transform (DFT) along the subcarrier axis is followed by depthwise and pointwise mixing and a forward DFT. The domain switch exposes delay-localized multipath structure directly. Unlike HybridDeepRx's pre-FFT waveform stage, the switch is internal to one block; the remaining blocks operate on the resource grid. The resulting architecture has 531,972 trainable parameters.

\begin{figure}[htbp]
\centering
\begin{tikzpicture}[
    font=\footnotesize,
    box/.style   = {draw, rounded corners=1pt, minimum height=6mm, align=center,
                    inner xsep=2pt, fill=black!3},
    nnbox/.style = {box, fill=blue!8, draw=blue!55},
    hyb/.style   = {box, fill=orange!16, draw=orange!75},
    sig/.style   = {align=center, inner sep=1pt},
    ar/.style    = {-{Latex[length=1.6mm,width=1.2mm]}, thick},
    cal/.style   = {dashed, gray!65, line width=0.3pt},
]

% ---------- Row 1: front end, then the network as a single block ----------
\node[sig]                     (rx)  {\(y(t)\)};
\node[box,   right=2mm of rx,    minimum width=12mm] (cp)  {Sync.,\\CP removal};
\node[box,   right=2mm of cp,    minimum width=12mm] (fft) {FFT};
\node[box,   right=3mm of fft,   minimum width=16mm] (gb)  {Remove\\guard, DC};
\node[nnbox, right=3.5mm of gb,  minimum width=20mm, minimum height=8mm] (nn)
      {\textbf{Hybrid CM}\\\textbf{DeepRx}};
\node[sig,   right=3mm of nn]  (llr) {LLRs};

\draw[ar] (rx)  -- (cp);
\draw[ar] (cp)  -- (fft);
\draw[ar] (fft) -- (gb);
\draw[ar] (gb)  -- node[above=-0.5pt, font=\footnotesize] {\(\mathbf{Y}\)} (nn);
\draw[ar] (nn)  -- (llr);

\node[sig, above=4mm of nn] (pil) {pilots \(x_{ij}\)};
\draw[ar] (pil) -- (nn);

% ---------- Row 2: inside the network ----------
\node[nnbox, below=12mm of cp.west, anchor=west, minimum width=17.5mm] (emb)  {Patch\\embed};
\node[hyb,   right=3mm of emb, minimum width=17.5mm] (hb)   {Hybrid block\\(delay domain)};
\node[nnbox, right=3mm of hb,  minimum width=17.5mm] (cm)   {\(5\times\) ConvMixer\\block};
\node[nnbox, right=3mm of cm,  minimum width=17.5mm] (head) {Output\\conv};

\draw[ar] (emb) -- (hb);
\draw[ar] (hb)  -- (cm);
\draw[ar] (cm)  -- (head);

\draw[cal] (nn.south west) -- (emb.north west);
\draw[cal] (nn.south east) -- (head.north east);

% ---------- Row 3: inside the hybrid block ----------
\node[hyb, below=10mm of emb.west, anchor=west, minimum width=17.5mm, minimum height=5.5mm] (ifft) {IFFT};
\node[hyb, right=3mm of ifft, minimum width=17.5mm, minimum height=5.5mm] (dw)   {Depthwise\\conv (delay)};
\node[hyb, right=3mm of dw,   minimum width=17.5mm, minimum height=5.5mm] (pw)   {Pointwise\\complex mix};
\node[hyb, right=3mm of pw,   minimum width=17.5mm, minimum height=5.5mm] (ffti) {FFT};

\draw[ar] (ifft) -- (dw);
\draw[ar] (dw)   -- (pw);
\draw[ar] (pw)   -- (ffti);

\draw[cal] (hb.south west) -- (ifft.north west);
\draw[cal] (hb.south east) -- (ffti.north east);

\end{tikzpicture}
\caption{Receive chain and Hybrid CM DeepRx architecture. Sync., CP, and conv denote synchronization, cyclic prefix, and convolution; FFT and IFFT denote the fast Fourier transform and its inverse.}
\label{fig:receiver}
\end{figure}
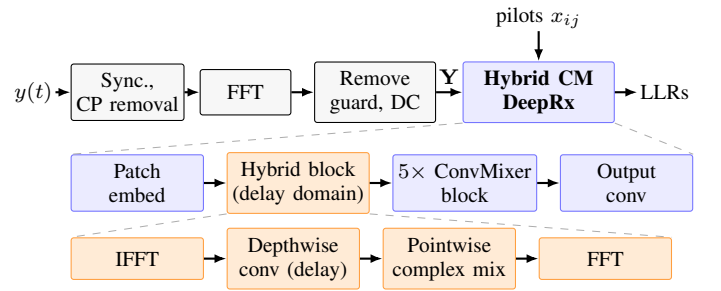

The pointwise stage uses complex multiplication and can represent phase rotations. Both mixing stages use pre-activation residual connections and omit batch normalization, following HybridDeepRx \cite{pihlajasalo_hybriddeeprx_2021}. Fig.~\ref{fig:receiver} summarizes the architecture and receive chain.

\subsection{Measurement Campaign}
\label{sec:measurement}

The dataset comes from the 5.88~GHz campaign of \cite{luostari_input-correlated_2026}, which used modified PlutoSDRs \cite{collins_software-defined_2018}. The receiver was mounted at base-station sites at heights of 4.5 to 7~m (Fig.~\ref{fig:hardware}); the transmitter was handheld or vehicle-mounted. An external amplifier provided up to 30~dBm transmit power, with antenna gains of approximately 5~dBi at the transmitter and 7~dBi at the receiver. The received signals reflect the actual propagation channel and receiver noise, together with practical impairments commonly omitted from synthetic channel models, such as power-amplifier nonlinearity, in-phase/quadrature (I/Q) imbalance, phase noise, and timing and frequency errors.

The eight sites span urban, suburban, rural, and industrial environments under pedestrian (0--3~m/s) and vehicular (up to 15~m/s) mobility. Five urban sites and one rural site were measured on foot; one suburban site was measured on foot and from a vehicle. The remaining industrial site contains scattered steel halls that produce strong reflections and was measured only from a vehicle. It is the only site of its category and the only entirely vehicular site. It was selected as the deployment site on physical grounds before model training or evaluation.

Slot synchronization, resource-grid demodulation, pilot-based signal-to-interference-plus-noise ratio (SINR) estimation, and cyclic redundancy check (CRC) filtering follow \cite{luostari_input-correlated_2026}. SINR uses the noise floor measured between transmissions. Only the preamble and pilots are known a priori. Payload-bit labels are obtained by decoding at code rate 0.25 after maximal-ratio combining (MRC) of both receive chains. Successful CRC validation makes these labels effectively error-free for training and bit error rate (BER) evaluation; fewer than 1\% of slots fail CRC and are discarded. The low code rate was chosen to support reliable label recovery. Although the SDR link can support other modulation and coding schemes, evaluating them would require additional OTA data; this study fixes the modulation and coding scheme and focuses on uncoded-BER sensitivity gains.

\subsection{Training Arms and Dataset Splits}
\label{sec:arms}

Unlike the pooled session-based splits in \cite{luostari_input-correlated_2026}, this experiment resplits the data by site, with slot rows randomly assigned within each site. Each row contains one transmission received through two spatially separated single-antenna chains; the chains are evaluated as separate SISO measurements but share the same transmission and location. Table~\ref{tab:splits} lists the resulting measurement counts. Only 1,000 deployment-site measurements enter training, and neither test set informs any training decision.

\begin{table}[htbp]
\caption{Dataset splits in measurements.}
\label{tab:splits}
\centering
\footnotesize
\renewcommand{\arraystretch}{0.75}
\begin{tabular*}{\columnwidth}{@{\extracolsep{\fill}}llr@{}}
\toprule
\textbf{Split} & \textbf{Contents} & \textbf{Count} \\
\midrule
Other-site training   & the seven other sites                    & 27,946 \\
All-site training     & the above plus the finetuning set        & 28,946 \\
Target-site finetuning & deployment site, for adaptation         & 1,000 \\
\midrule
Target-site test      & deployment site, held out                & 2,488 \\
Other-site test       & other sites, held out                    & 1,000 \\
Validation            & other sites                              & 800 \\
\bottomrule
\end{tabular*}
\renewcommand{\arraystretch}{1}
\end{table}

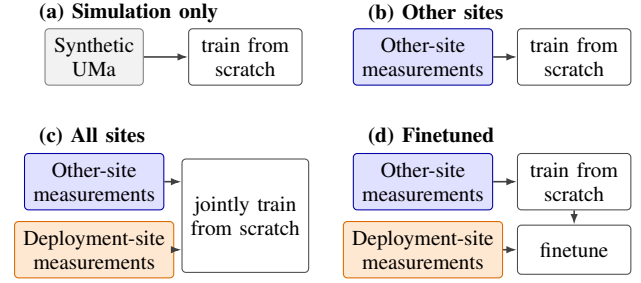
\begin{figure}[htbp]
\centering
\begin{tikzpicture}[
    font=\footnotesize,
    x=1cm, y=1cm,
    sim/.style={draw=black!50, fill=black!5, rounded corners=1.5pt,
                minimum width=13mm, minimum height=6.5mm, align=center},
    other/.style={draw=blue!65!black, fill=blue!12, rounded corners=1.5pt,
                  minimum width=17mm, minimum height=6.5mm, align=center},
    target/.style={draw=orange!85!black, fill=orange!18, rounded corners=1.5pt,
                   minimum width=17mm, minimum height=6.5mm, align=center},
    operation/.style={draw=black!70, fill=white, rounded corners=1.5pt,
                      minimum width=15mm, minimum height=6.5mm, align=center},
    arm/.style={font=\footnotesize\bfseries, anchor=west},
    flow/.style={-{Latex[length=1.4mm,width=1mm]}, semithick, draw=black!75},
]

\node[arm] at (0,0.58) {(a) Simulation only};
\node[sim] (sdata) at (0.85,0) {Synthetic\\UMa};
\node[operation] (strain) at (2.85,0) {train from\\scratch};
\draw[flow] (sdata) -- (strain);

\node[arm] at (4.35,0.58) {(b) Other sites};
\node[other] (odata) at (5.20,0) {Other-site\\measurements};
\node[operation] (otrain) at (7.20,0) {train from\\scratch};
\draw[flow] (odata) -- (otrain);

\node[arm] at (0,-1.05) {(c) All sites};
\node[other] (aother) at (0.85,-1.65) {Other-site\\measurements};
\node[target] (atarget) at (0.85,-2.55) {Deployment-site\\measurements};
\node[operation, minimum height=15mm] (ajoint) at (2.85,-2.10) {jointly train\\from scratch};
\draw[flow] (aother.east) -- (aother.east -| ajoint.west);
\draw[flow] (atarget.east) -- (atarget.east -| ajoint.west);

\node[arm] at (4.35,-1.05) {(d) Finetuned};
\node[other] (fother) at (5.20,-1.65) {Other-site\\measurements};
\node[operation] (ftrain) at (7.20,-1.65) {train from\\scratch};
\node[target] (ftarget) at (5.20,-2.55) {Deployment-site\\measurements};
\node[operation] (ftune) at (7.20,-2.55) {finetune};
\draw[flow] (fother) -- (ftrain);
\draw[flow] (ftarget) -- (ftune);
\draw[flow] (ftrain.south) -- (ftune.north);

\end{tikzpicture}
\caption{Training data and optimization flow for the four arms. UMa denotes urban macro.}
\label{fig:arms}
\end{figure}

Fig.~\ref{fig:arms} summarizes the four training arms.

All four arms use the same receiver, described in Section~\ref{sec:deeprx}. Other-site versus simulation measures cross-site transfer, all-site versus other-site measures the value of deployment-site data, and finetuned versus all-site compares optimization protocols using identical measurements. The \emph{simulation} arm trains only on synthetic Third Generation Partnership Project (3GPP) urban macro (UMa) channels generated with Sionna \cite{3GPP:38.901,hoydis_sionna_2023}; the \emph{other-site} arm uses the seven other sites; the \emph{all-site} arm adds the deployment-site data jointly; and the \emph{finetuned} arm continues from the other-site model using only deployment-site data.

All arms use AdamW and an SINR-weighted binary cross-entropy loss. The three from-scratch arms train for approximately 40,000 iterations at batch size 32.

For data augmentation, the measured-data arms add AWGN that lowers SINR by a uniformly sampled amount up to 1~dB. Simulated samples are not augmented because each one is generated with a freshly drawn channel, noise realization, and SINR.

The all-site and finetuned arms use the same 28,946 measurements. All-site training mixes deployment-site data throughout training from scratch, whereas finetuning concentrates it in a second stage with its own budget and settings (Section~\ref{sec:finetune}). The finetuned model starts from the other-site checkpoint, preserving other-site exposure without seeing deployment-site data twice.

\subsection{Site-Specific Finetuning}
\label{sec:finetune}

Finetuning uses deployment-site data exclusively for 10,000 iterations at batch size 16 and a fixed learning rate of \(4\times10^{-6}\), with no warmup or decay. The adaptation-set size of 1,000 measurements and the fixed learning rate were selected through auxiliary exploration without using either test set. There is no early stopping or checkpoint selection; evaluation uses the final checkpoint, so held-out data do not influence training. To trace the adaptation trajectory, the finetuning stage was rerun for each seed with checkpoints saved every 500 iterations (Fig.~\ref{fig:curve}).

\subsection{Evaluation Metrics}
\label{sec:metrics}

Sensitivity gain is the decrease in SINR required to reach a target uncoded BER, obtained by logarithmic interpolation of the binned BER curves. Each test set's LS RX crossing serves as the local reference, so models are compared with LS RX on the same test set rather than across test sets.

We evaluate two targets, selected from auxiliary link-level simulations with the same numerology and 16QAM. An uncoded BER of \(10^{-1}\) corresponds approximately to a 10\% coded block error rate (BLER) at code rate 0.5, while \(10^{-2}\) is representative of code rates near 0.9. The two targets therefore represent moderate- and higher-rate coded links. Both provide a common crossing for all evaluated receivers within the measured SINR window. We identify the operating point in every transfer claim.

\subsection{Replicates and Statistical Analysis}
\label{sec:setup}

Each arm is trained under five independent seeds, each defining a full pipeline replicate. For each seed, the finetuned model continues from the corresponding other-site model, so arm comparisons are paired by seed. We compute each difference within seed and report its mean and standard deviation across the five seeds rather than subtracting arm means. Displayed arm means are rounded independently, whereas paired differences are computed from unrounded per-seed values.

\section{Results}

\subsection{Transfer of Over-the-Air Data Across Sites}
\label{sec:transfer}

We assess cross-site transfer in two steps. First, on the deployment site, the other-site model provides \(0.29 \pm 0.03\)~dB more sensitivity gain than the simulation-only model at \(10^{-1}\) and \(0.12 \pm 0.05\)~dB more at \(10^{-2}\) (Table~\ref{tab:target}), showing beneficial transfer at both operating points.

\begin{table}[htbp]
\caption{Deployment-site sensitivity gain over LS RX, reported as five-seed mean \(\pm\) across-seed standard deviation. LS RX crosses \(10^{-1}\) and \(10^{-2}\) at 9.65 and 17.95~dB, respectively.}
\label{tab:target}
\centering
\footnotesize
\renewcommand{\arraystretch}{0.75}
\begin{tabular*}{\columnwidth}{@{\extracolsep{\fill}}lcc@{}}
\toprule
\textbf{Training data} & \textbf{Gain at \(10^{-1}\)} & \textbf{Gain at \(10^{-2}\)} \\
\midrule
Simulation only  & \(1.62 \pm 0.02\)~dB   & \(2.32 \pm 0.04\)~dB \\
Other sites      & \(1.91 \pm 0.02\)~dB   & \(2.44 \pm 0.05\)~dB \\
All sites        & \(1.94 \pm 0.01\)~dB   & \(2.73 \pm 0.04\)~dB \\
Finetuned        & \(\mathbf{1.97 \pm 0.01}\)~\textbf{dB} & \(\mathbf{2.91 \pm 0.04}\)~\textbf{dB} \\
\bottomrule
\end{tabular*}
\renewcommand{\arraystretch}{1}
\end{table}

Second, we compare the same other-site model across test sets. Its sensitivity gain is \(2.26 \pm 0.01\)~dB on the other-site test set and \(1.91 \pm 0.02\)~dB at the deployment site at \(10^{-1}\), and \(2.58 \pm 0.02\) and \(2.44 \pm 0.05\)~dB at \(10^{-2}\) (Tables~\ref{tab:target} and \ref{tab:others}). The unseen-site gap is therefore 0.35~dB at \(10^{-1}\) and 0.14~dB at \(10^{-2}\). The test sets have nearly identical SINR distributions, so a simple SINR-distribution shift does not explain this gap. Comparing each model against the local LS RX reduces but does not eliminate other differences in test-set difficulty.

\begin{table}[htbp]
\caption{Other-site sensitivity gain over LS RX, reported as five-seed mean \(\pm\) across-seed standard deviation. LS RX crosses \(10^{-1}\) and \(10^{-2}\) at 9.99 and 17.72~dB, respectively.}
\label{tab:others}
\centering
\footnotesize
\renewcommand{\arraystretch}{0.75}
\begin{tabular*}{\columnwidth}{@{\extracolsep{\fill}}lcc@{}}
\toprule
\textbf{Training data} & \textbf{Gain at \(10^{-1}\)} & \textbf{Gain at \(10^{-2}\)} \\
\midrule
Simulation only  & \(1.82 \pm 0.01\)~dB   & \(2.08 \pm 0.03\)~dB \\
Other sites      & \(\mathbf{2.26 \pm 0.01}\)~\textbf{dB} & \(\mathbf{2.58 \pm 0.02}\)~\textbf{dB} \\
All sites        & \(\mathbf{2.26 \pm 0.01}\)~\textbf{dB} & \(\mathbf{2.57 \pm 0.02}\)~\textbf{dB} \\
Finetuned        & \(2.22 \pm 0.02\)~dB   & \(2.47 \pm 0.03\)~dB \\
\bottomrule
\end{tabular*}
\renewcommand{\arraystretch}{1}
\end{table}

\subsection{Value of Deployment-Site Data}
\label{sec:necessity}

\begin{figure}[htbp]
  \centering
  \includegraphics[width=\columnwidth]{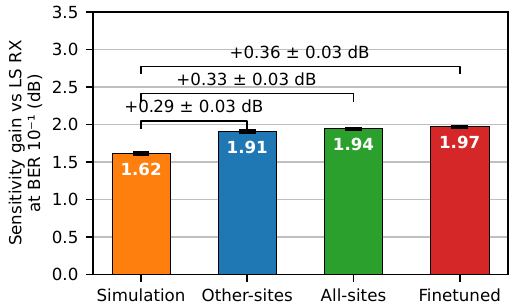}
  \caption{Deployment-site sensitivity gain over LS RX at an uncoded BER of \(10^{-1}\). Bars show five-seed mean total gains, error bars show across-seed standard deviations, and brackets show paired differences from simulation. LS RX reaches \(10^{-1}\) at 9.65~dB.}
  \label{fig:seedbarslow}
  \end{figure}

\begin{figure}[htbp]
\centering
\includegraphics[width=\columnwidth]{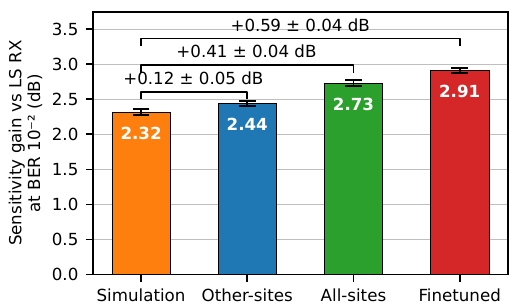}
\caption{Deployment-site sensitivity gain over LS RX at an uncoded BER of \(10^{-2}\), shown as in Fig.~\ref{fig:seedbarslow}. LS RX reaches \(10^{-2}\) at 17.95~dB.}
\label{fig:seedbars}
\end{figure}

To measure the value of deployment-site data, we compare joint training and finetuning with the other-site model. At \(10^{-1}\), joint training improves deployment-site sensitivity by \(0.04 \pm 0.01\)~dB and finetuning by \(0.07 \pm 0.01\)~dB, for total gains over simulation of \(0.33 \pm 0.03\) and \(0.36 \pm 0.03\)~dB, respectively (Fig.~\ref{fig:seedbarslow}). At \(10^{-2}\), the corresponding improvements are \(0.29 \pm 0.06\) and \(0.47 \pm 0.01\)~dB, for total gains of \(0.41 \pm 0.04\) and \(0.59 \pm 0.04\)~dB (Fig.~\ref{fig:seedbars}). Arm ordering is identical in every seed at both operating points. Deployment-site data therefore improves sensitivity at both operating points, with the larger increment at \(10^{-2}\).

\begin{figure}[htbp]
\centering
\includegraphics[width=\columnwidth]{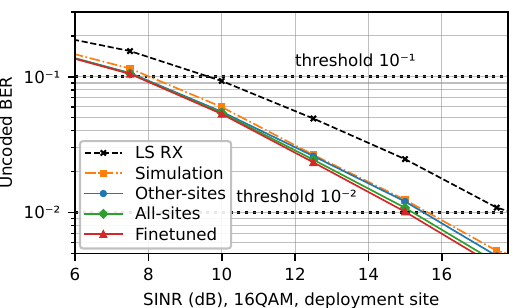}
\caption{Five-seed mean uncoded BER versus SINR on the deployment site for the four training arms and the LS RX baseline, shown over 6--18~dB to span both sensitivity thresholds.}
\label{fig:bervssinr}
\end{figure}

The all-site and finetuned models use identical measurements but incorporate them differently during optimization. Relative to joint training, finetuning adds 0.03~dB of sensitivity gain at \(10^{-1}\) (across-seed standard deviation below 0.01~dB) and \(0.18 \pm 0.05\)~dB at \(10^{-2}\) (Fig.~\ref{fig:bervssinr}).

\subsection{Adaptation versus Forgetting}
\label{sec:forgetting}

We assess spatial forgetting on the other-site test set. The paired sensitivity difference between the jointly trained and other-site models (joint minus other-site) is \(0.00 \pm 0.01\)~dB at \(10^{-1}\) and \(-0.01 \pm 0.02\)~dB at \(10^{-2}\) (Table~\ref{tab:others}). Thus, joint training adds deployment-site data with no measurable aggregate loss on the other-site test set.

With 1,000 deployment-site measurements and a fixed learning rate of \(4\times10^{-6}\), finetuning improves deployment-site sensitivity but lowers sensitivity gain at the other sites by \(0.04 \pm 0.01\)~dB at \(10^{-1}\) and \(0.11 \pm 0.04\)~dB at \(10^{-2}\) relative to the other-site starting model. We call this measured loss spatial forgetting. Even after this loss, the finetuned model retains about 0.4~dB more sensitivity gain than the simulation-only model on the aggregate other-site test set at both operating points (Table~\ref{tab:others}).

\begin{figure}[htbp]
  \centering
  \includegraphics[width=\columnwidth]{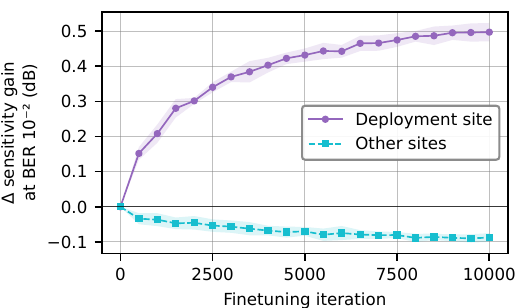}
  \caption{Adaptation and forgetting during finetuning: change in sensitivity gain at an uncoded BER of \(10^{-2}\) relative to each run's starting other-site model, on both test sets. Curves average five repeated finetuning runs, one per seed; shading denotes \(\pm 1\) across-seed standard deviation. Positive values indicate improvement.}
  \label{fig:curve}
  \end{figure}

Fig.~\ref{fig:curve} scores the repeated finetuning runs on both test sets at \(10^{-2}\). At 10,000 iterations, they reach \(0.50 \pm 0.03\)~dB at the deployment site and \(-0.09 \pm 0.01\)~dB at the other sites. These values differ slightly from the paired differences above (0.47 and \(-0.11\)~dB) because they come from separate finetuning runs. By 5,000 iterations, deployment-site sensitivity has improved by \(0.43 \pm 0.02\)~dB and other-site sensitivity has decreased by \(0.07 \pm 0.01\)~dB. The final 5,000 iterations add about 0.07~dB at the deployment site and remove about 0.02~dB at the other sites, and both curves flatten toward the end of the budget. The prescribed 10,000-iteration budget was not selected as an optimum.

\balance

\section{Discussion and Limitations}
\label{sec:discussion}

Other-site data transfers at both operating points, whereas deployment-site data adds little at \(10^{-1}\) but substantially more at \(10^{-2}\). At both points, finetuning yields greater deployment-site sensitivity than joint training. The OTA-trained arms lie within 0.07~dB of each other at the \(10^{-1}\) crossing and separate toward the \(10^{-2}\) crossing, while the simulation-only model trails them at both (Fig.~\ref{fig:bervssinr}). The relative importance of data origin and adaptation protocol therefore depends on the operating point. A plausible explanation is that at low SINR, errors are dominated by noise, which limits all receivers alike, whereas at higher SINR residual errors increasingly stem from channel-estimation mismatch, where site-specific propagation and mobility matter more.

Because every site used the same SDR transmitter and receiver and the same synchronization and demodulation chain, OTA-trained models may also learn common hardware impairments and residual timing or frequency errors. This shared contribution is part of the other-site increment at the deployment site, 0.12~dB at \(10^{-2}\), whereas deployment-site data added a further 0.29 to 0.47~dB there. At the higher-rate operating point, local propagation and mobility therefore appear to account for most of the gain; controlled ablations are needed to separate these effects fully.

The two adaptation protocols differ in training budget, batch size, and data exposure. Deployment-site measurements constitute 3.5\% of the all-site corpus, and over the full budgets each one is presented about 160 times during finetuning and about 45 times during joint training. The experiment therefore compares complete protocols rather than isolating a single cause, and protocol performance may depend on these proportions. Two observations indicate that the finetuning gain reflects the deployment-site data rather than additional optimization. First, finetuning raised deployment-site sensitivity while lowering other-site sensitivity (Fig.~\ref{fig:curve}). Second, in exploratory runs over several seeds, continuing other-site training at a lower learning rate yielded no measurable gain. In an exploratory single-seed experiment, increasing the deployment-site subset from 1,000 to 1,900 measurements yielded only a small additional gain at \(10^{-2}\), suggesting diminishing returns beyond 1,000 measurements.

In practice, simulation-only training is a fallback when OTA measurements are unavailable, and other-site measurements provide a transferable starting point before deployment-site data are collected. Under the tested protocols, joint training shows no measurable aggregate loss on the other-site test set, whereas finetuning provides the larger local gain at the cost of spatial forgetting. Joint training keeps the source sites in every batch, which acts as rehearsal and is consistent with its lack of measurable forgetting, whereas finetuning on deployment-site data alone removes this constraint. Baytekin et al. found that finetuning gains carry over to other scenarios relative to a synthetically pretrained receiver \cite{baytekin_site-specific_2026}. Our results agree when measured against simulation-only training, since the finetuned model remains about 0.4~dB ahead at the other sites, but relative to an OTA-trained starting point, adapting to one site costs sensitivity elsewhere.

Several limitations qualify these findings. The single held-out site is both the only industrial site and the only entirely vehicular one, so site identity is confounded with environment and mobility; leave-one-site-out evaluation is the natural next step. The splits are random rather than route- or time-blocked, so finetuning and test measurements at the deployment site may share drive routes, and part of the deployment-site gain may reflect location-specific rather than site-wide adaptation.

The quantitative magnitudes are specific to the single Hybrid CM DeepRx architecture, this campaign's hardware, the SISO 5.88~GHz 16QAM link, measured environments and mobility, dataset size, evaluated protocols, and operating points. Because deployment-site data contributed most at the higher-rate operating point, its value may increase for higher-order modulations such as 64QAM and 256QAM, which operate at higher SINR. Multi-antenna links would additionally expose site-specific spatial correlation that a SISO link cannot capture. Finally, no deployment-only model was trained from scratch. With a deployment-site corpus comparable in size to the other-site corpus, such a model may achieve the highest deployment-site sensitivity and would provide an upper bound for transfer and finetuning; this is left for future work.

\section{Conclusion}

This eight-site study shows that OTA measurements from seven source sites transfer to a distinct held-out deployment site. The additional value of deployment-site data depends on the operating point: it is small at the moderate-rate point but substantial at the higher-rate point. Because the hardware is shared across sites, this higher-rate gain points mainly to local propagation and mobility. Joint training incorporates deployment-site data without measurable aggregate loss on the other-site test set, whereas finetuning yields the larger local gain at the cost of about 0.1~dB of spatial forgetting at the higher-rate point; even then, it remains about 0.4~dB ahead of simulation-only training on the aggregate other-site test set. Although based on a single held-out site, these results suggest a practical hierarchy: use simulation when OTA data are unavailable, use other-site measurements as a transferable starting point, and add deployment-site data especially when higher-rate sensitivity matters. Under the tested protocols, joint training is preferable when aggregate other-site retention is the priority, and finetuning when deployment-site sensitivity justifies a small loss elsewhere.

\section*{Acknowledgment}

The authors thank Mr. Geoff Tunnicliffe for facilitating the experimental transmit license for the measurement campaign.

\footnotesize
\bibliographystyle{IEEEtran}
\bibliography{SiteSpecific_OTA,bstctl}

\end{document}